\documentclass[aps,prl,longbibliography,superscriptaddress,twocolumn,10pt,nofootinbib]{revtex4-1}
\usepackage[utf8]{inputenc}
\usepackage[english]{babel}
\usepackage{times}
\usepackage{color}
\usepackage{graphicx}
\usepackage{import}
\usepackage{amsmath}
\usepackage{braket}

\usepackage{hyperref}
\usepackage[all]{hypcap}

\usepackage{booktabs}

\AtBeginDocument{
\heavyrulewidth=.08em
\lightrulewidth=.05em
\cmidrulewidth=.03em
\belowrulesep=.65ex
\belowbottomsep=0pt
\aboverulesep=.4ex
\abovetopsep=0pt
\cmidrulesep=\doublerulesep{}
\cmidrulekern=.5em
\defaultaddspace=.5em
}

\begin{document}
\title{Optimization of Lattice Surgery is Hard}
\author{Daniel Herr}
\email{daniel.herr@riken.jp}
\affiliation{Quantum Condensed Matter Research Group, CEMS, RIKEN, Wako-shi 351-0198, Japan}
\affiliation{Computational Physics, ETH Zurich, 8093 Zurich, Switzerland}
\author{Franco Nori}
\affiliation{Quantum Condensed Matter Research Group, CEMS, RIKEN, Wako-shi 351-0198, Japan}
\affiliation{Department of Physics, University of Michigan, Ann Arbor, MI 48109-1040, USA}
\author{Simon J. Devitt}
\affiliation{ARC Centre for Engineered Quantum Systems, Department of Physics and Astronomy, Macquarie University, North Ryde, New South Wales 2109, Australia}

\begin{abstract}
The traditional method for computation in either the surface code or in the Raussendorf model is the creation of holes or ``defects'' within the encoded lattice of qubits that are manipulated via topological braiding to enact logic gates. However, this is not the only way to achieve universal, fault-tolerant computation.
In this work, we focus on the Lattice Surgery representation, which realizes transversal logic operations without destroying the intrinsic 2D nearest-neighbor properties of the braid-based surface code and achieves universality without defects and braid based logic.
For both techniques there are open questions regarding the compilation and resource optimization of quantum circuits. Optimization in braid-based logic is proving to be difficult and the classical complexity associated with this problem has yet to be determined. In the context of lattice-surgery-based logic, we can introduce an optimality condition, which corresponds to a circuit with the lowest resource requirements in terms of physical qubits and computational time, and prove that the complexity of optimizing a quantum circuit in the lattice surgery model is NP-hard.
\end{abstract}

\maketitle

\section{Introduction}
  Quantum computation is a very promising method to perform information processing. Several types of problems, such as prime factorization~\cite{shor_factor} or search algorithms~\cite{grover_search} can be sped up considerably. The first physical realizations have been built~\cite{superconducting_err_cor,VD14,TT16,L16,LM16}, where error rates are small enough to allow for effective error-correction and fault-tolerant quantum computation~\cite{Fowler2012}.
  Many diverse systems can already run small quantum algorithms and projects such as the IBM Quantum Experience~\cite{IBM_quantum}, have connected small prototype computers to the cloud for both educational purposes and to allow other researchers to test small scale protocols.
  
  One of the remaining tasks for experimentalists is to scale up the number of qubits, while maintaining 
  low error rates, in order to allow more complex algorithms to be performed. The task for theorists is now to build a quantum compiler~\cite{liquid,haener2016,PPND15,Braiding_compiler}, that can translate high-level algorithms to individual hardware instructions. This compiler has to be aware of the hardware faults and should introduce error-correction to protect logical qubits from physical influences, in order to ensure a completely fault-tolerant computation. Several parts of such a compiler have already been created~\cite{liquid,haener2016,PPND15,Braiding_compiler}, but a complete software package has yet to be developed. Notably, optimization algorithms~\cite{PF13} which operate at the error-correction level are still lacking to optimize physical resources in the most commonly used error-correction models. To this end, we inspect the optimization of a specific topologically based operational model called Lattice Surgery (LS)~\cite{Horsman2012}. This representation was chosen particularly because of its applicability to a wide range of hardware models~\cite{JMFMKLY10,NTDS13,HPHHFRSH15,LHMB15,LWFMDWH15}, and the applicability of LS approaches using other topological coding techniques~\cite{T15,DIP16,B16}.

  For a practical fault-tolerant computer using LS, both the physical and logical level is arranged in a 2D nearest-neighbor array, on which a universal gate set can be realized~\cite{Horsman2012,herr_lattice_surgery}. This 2D, nearest-neighbor environment is enforced by the connectivity of the physical qubit array. For LS this is the planar code~\cite{DKLP02}, for braiding it is generally the surface code~\cite{DKLP02,FSG08,Fowler2012}.
  The common feature of all these representations is that physical qubits are connected via a graph, that indicates their possible interactions. Even non-fault-tolerant implementations suffer from the restricted connectiveness of the underlying physical qubits and methods to perform computation on these had to be developed~\cite{distr_quantum}. 

  Conceptually, algorithmic compilation and optimization is similar to more traditional measurement-based quantum computation, but at the level of error-corrected qubits.  The LS translation~\cite{herr_lattice_surgery} of an arbitrary circuit creates an algorithmically specific graph state at the encoded level, using the native parity checks of LS.  After this encoded graph is created, a time-ordered sequence of non-Clifford measurements is performed on each encoded node in the graph to realize the algorithm.  This is akin to traditional measurement based quantum computation~\cite{RBB03} (which is not error-corrected), where a 2D, universal graph state (commonly referred to as a cluster state) is prepared, a quantum circuit mapped to this 2D array and all associated Clifford measurements are performed.  The 2D cluster state is then converted to an algorithmically specific graph state where the only subsequent operations needed are a time-ordered sequence of non-Clifford measurements and feedforward~\cite{F12,herr_lattice_surgery}.  
  
  Here, we want to evaluate the complexity of the creation of such an encoded graph state. In complexity theory, problems are divided into categories, which determine their hardness. A famous class consists of nondeterministic polynomial complete (NP-complete) problems~\cite{Moore2011}. Such problems lie in the complexity class NP, such that a solution can be verified in polynomial time, and are at least as hard as the most difficult problems in NP. A common way to determine NP-completeness is to map an already known NP-complete problem to the problem of interest~\cite{Moore2011}. We were inspired by the proof of NP-hardness of Tetris~\cite{tetris}, and map the 3-partitioning problem~\cite{NP-compBook} to the optimization of LS patches using the translation devised in~\cite{herr_lattice_surgery}. This implies that it is also NP-hard to optimize the complete problem including measurements, because this only adds an additional layer of complexity to the system.

  We will proof that the circuit optimization of a particular fault-tolerant implementation of topological error-correction is NP-hard. Similarly to our result, it has been shown that it can be NP-hard for a compiler to optimize classical code, such that its execution is time optimal~\cite{class_optimization}. Our results, thus, urge the development of heuristics that can optimize quantum circuits not exactly, but at least reasonably well, for implementation on realistic quantum hardware. Furthermore, we derive general estimates on best and worst performance. We also discuss the benefit of optimization given a sample algorithm and estimate the hardness of the optimization problem for an exact, classical solver.

\subsection{Revision of Lattice Surgery}
  The main idea of the LS translation is to encode an algorithmically-specific graph state in the square lattice of the planar code, which will then use a measurement-based quantum computational approach to perform any calculation. The implementation of this encoded state needs to respect the underlying structure of the planar code. Many square patches which encode individual qubits~\cite{Horsman2012} are aligned on a 2D lattice. Connections between nearest-neighbor logical qubits are possible using physical qubits that lie on the boundary between the patches.
  These operations constitute merges and splits, that act as parity checks between the two encoded qubits, and can be used together with injection to enable universal quantum computation~\cite{herr_lattice_surgery}.

  The analysis performed here is rooted in the LS translation given in~\cite{herr_lattice_surgery}. First, patches are initialized to $\ket{+}$; then, using parity checks, a algorithmically specific stabiliser state is generated. This stabiliser state is measured in the bases $\ket{Z}$,$\ket{X}$,$\ket{Y} = P\ket{+}$ and $\ket{A} = T\ket{+}$, where $P = \sqrt{Z}$ and $T=\sqrt{P}$. However, for planar codes the rotated basis measurements ($\ket{Y}$, $\ket{A}$) are not protected fault-tolerantly, and magic states must be injected~\cite{Horsman2012,herr_lattice_surgery}. Our description is only concerned with the creation of the initial algorithmically specific stabiliser state and shows that even the optimization of this less complicated problem is already NP-hard.

\begin{figure}
    \centering
    \includegraphics[width=\columnwidth]{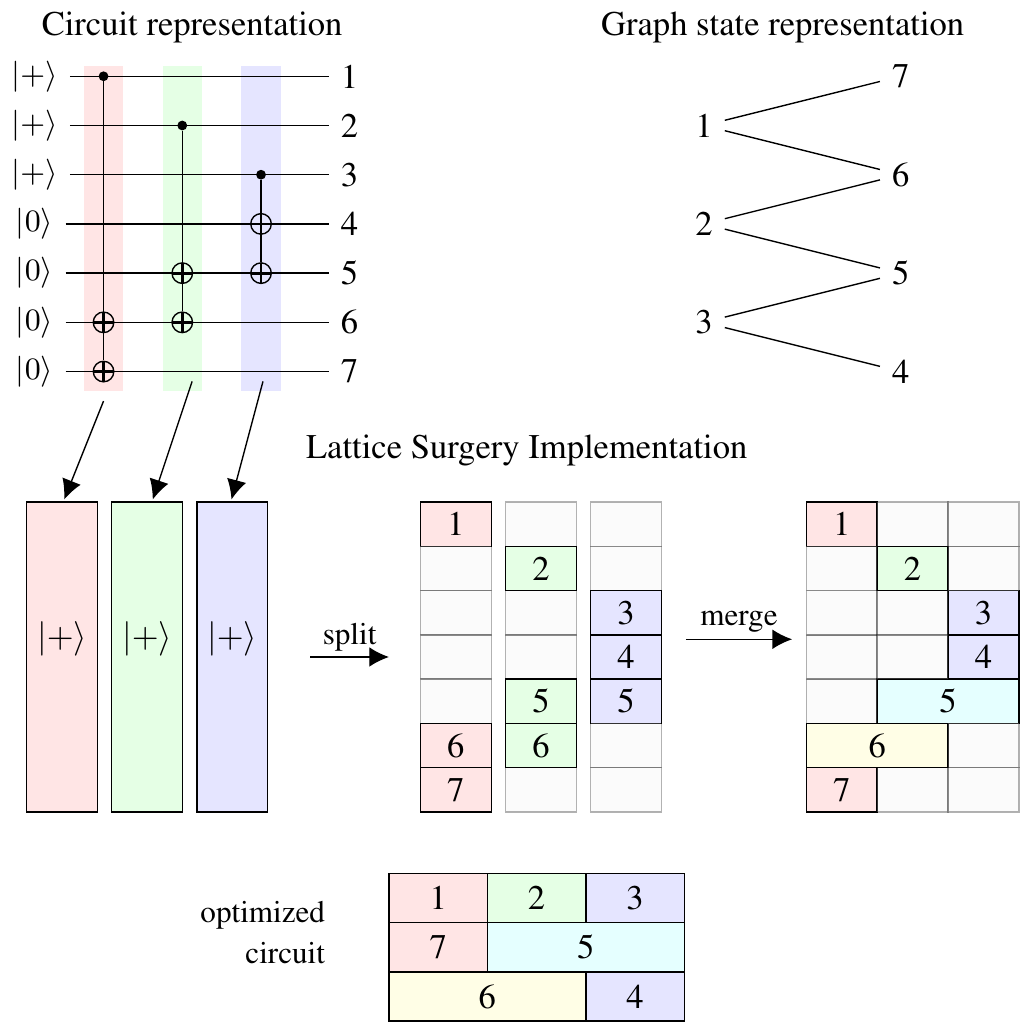}
    \caption{\label{fig:cnot_split} Here multi-target CNOTs are implemented using LS. During initialization three patches of surface code with $N$ by $7 N$ qubits are created, which are then split and merged to perform the computation. The faded boxes indicate ancillary qubits. An optimized version of this circuit is shown at the bottom where the placement of the patches ensures a minimal bounding box of the whole circuit area. This circuit can achieve the theoretical optimality.}
  \end{figure}

  An arbitrary circuit can be rearranged into the ICM format~\cite{PPND15}, which is already divided into (I)nitializations, (C)NOTs and (M)easurements. The first two steps can be interpreted as a circuit to generate the stabiliser state. The translation to LS first merges all CNOTs of this circuit into multi-target CNOTs, which can then be easily implemented in LS: For each multi-target CNOT a column in the planar code is created, which is later split into individual encoded qubits (Figure~\ref{fig:cnot_split}). Then, the qubits which are targeted by two or more CNOTs have to be combined through LS merge operations.
  
  Due to different CNOTs targeting the same qubits multiple times in a general quantum circuit, the compiler naturally produces ancillary qubits, which are inherent to the structure of the algorithm and the compiler. During our calculations these are disregarded and we only study the problem of how to optimally place the patches in the 2D-nearest neighbor environment of the planar code error-correction model using LS.

  \section{Optimality}
  A usual definition of optimality is reaching a (computational) goal with minimal physical requirements. In our case these physical requirements correspond to a minimal space-time volume, which is defined by the product of error-correcting cycles and physical qubits.
  We will further restrict this definition such that the bounding box of this space-time volume (within which all computation happens) needs to be minimal, while the placement still retains the same output state. Another way at looking at this definition is that every patch of the surface code inside the bounding box is initialized to a computational qubit and no ancillary patches are needed.
  We focus on the generation of the algorithmically-specific stabilizer-state and prove that even the optimization of this part is NP-hard. Such a stabilizer-state can be prepared in constant time (as all circuit elements are Clifford), which allows a simplified optimality condition to be the mapping that results in the ``least surface area''.

\subsection{(Non-physical) Problem Description}

  The LS translation creates a problem where each CNOT has to be fitted into a surface code area that contains all computations. This area should be minimized. However, this can be viewed as an abstract problem, completely detached from the LS picture. We will now introduce the problem that needs to be solved.

  The problem consists of minimizing the surface area of a square lattice which consists of individual patches. Some of these patches are assigned an integer $q_{ij}$ but they do not necessarily need one. Furthermore, multiple patches can have the same integer.
  A horizontal (vertical) neighbor of a patch is defined as the next nonempty patch (i.e. patch that contains an integer) to the right or left (up or down).
  A set of boxes $C_i$ containing patches with integers $q_{ij}$ are given and can be implemented on the lattice by a chain of vertical neighbors, where the order of the $\{q_{ij}\}_j$ can be chosen freely. Furthermore empty patches can be added freely.
  The following criteria have to be met to obtain a valid configuration:
  \begin{enumerate}
  \item Patches for all boxes need to be placed (vertical neighbors);
  \item Patches with the same integers need to be placed such that they are horizontal neighbors.
  \end{enumerate}
  Thus, the problem consists of an optimal placement of these numbered patches such that the area of the bounding box of the total arrangement is minimized. The less empty patches are required the more optimized is the configuration.

  For a circuit that has been prepared in the universal, inverted ICM-representation each multi target CNOT operation will contribute to one box $C_i$. The numbers $q_{ij}$ of box $C_i$ are given by the qubits that partake in this operation. Figure~\ref{fig:cnot_split} shows how a sample circuit is mapped to the LS representation.

  If a circuit reaches optimality, the number of patches needed in LS can be calculated by:
  \begin{equation}
    N_\text{Patch} = \sum_{i = 1}^{\#\text{CNOT}} \left(N_{\text{target}_i} + 1 \right).
    \label{eq:cost}
  \end{equation}
  Here, $\#\text{CNOT}$ denotes the number of multi-target CNOTs with different qubits as their control, in the original circuit specification, and $N_{\text{target}_i}$ is the number of target qubits for the $i{\text{th}}$ CNOT.
  However, due to incompatibility during the merge step, this can (in the worst case) lead to a non-optimal placement with a patch-requirement of
  \begin{equation}
  N_\text{Patch} = N_Q \cdot \#\text{CNOT},
  \label{eq:cost_unopt}
  \end{equation}
  where $N_Q$ represents the total number of qubits in the circuit.

  Due to the structure of the high-level circuit that needs to be compiled, it is not always possible to reach the theoretical optimum. A general optimized algorithm needs resources between the two bounds given above.

\section{Proof of NP-completeness}
  We will now prove the NP-completeness of the decision problem of determining, whether the theoretical optimum can be reached.

  Inspired by the proof of NP-hardness of Tetris~\cite{tetris}, we will map the 3-partition problem to a circuit, which gets translated to LS. We will show, that with polynomial overhead a solution to the number partitioning problem can be obtained by the optimization of the placement of LS patches.

  In the 3-partitioning problem~\cite{NP-compBook}, a set of non-negative integers $\left\{a_i \right\}_{1\le i \le 3s}$ is given. With another non-negative integer $L$, two further requirements are: (i) $\frac{L}{4} \le a_i \le \frac{L}{3}$ $\forall i$ such that $1 \le i \le 3 s$ and (ii) $\sum_{i=1}^{3s} a_i = s L$.

  The NP-complete decision problem for 3-partitioning answers the following question: Can $\left\{a_i\right\}_{1 \le i \le 3s}$ be partitioned into $s$ disjoint subsets $A_1, \cdots , A_s$, such that $\sum_{i \in A_j} a_i = L$ for $j \in \left\{1,\cdots , s\right\}$?

\subsection{Mapping}
  \begin{figure}
    \centering
    \includegraphics[width=\columnwidth]{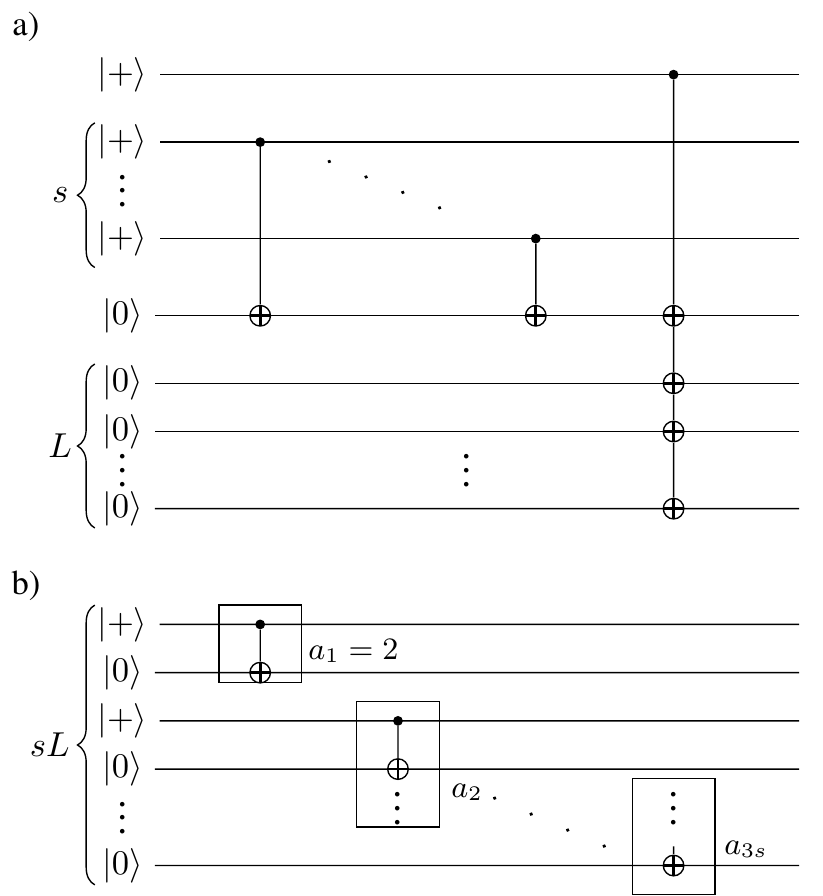}
    \caption{\label{fig:MappingCirc} The optimization for both parts of this circuit corresponds to solving the 3-partitioning problem.
    Part a) of this Figure implements the 3-partitioning problem only. Each CNOT corresponds to a number $a_i$ and will be translated into a separate patch of variable height in the LS representation of Figure~\ref{fig:MappingLS}. We call the area in the lattice surgery representation, which only consists of the CNOTs from part a), the \emph{compute area}.
    Part b) of this circuit is used to force the compute area to be a rectangle of height $L$ and width $s$ and the qubits in the compute area are responsible of encoding the original NP-complete problem.}
  \end{figure}
  We can translate the problem of 3-partitioning to the problem of deciding whether a corresponding circuit can reach optimality in LS. The main idea of this mapping is to encode each of the integers of the 3-partitioning problem $a_i$ into a single multi-target CNOT, where the number of qubits that partake in the $i$-th CNOT is given by $a_i$. Therefor, a box $C_i$ of the non-physical problem description contains $a_i$ integers which will then be translated to blocks of width 1 and height $a_i$ in the LS model, for optimality.
  Furthermore, each qubit is only acted on by one CNOT, such that no further constraints apply to the placement of these boxes.
  The solution of the 3-partitioning problem is given by finding an arrangement of these CNOT blocks in a rectangle of height $L$ and width $s$. We will call this rectangle the compute area.
  In Figure~\ref{fig:MappingCirc} we show a possible circuit, where the qubits in part a) implement the CNOTs corresponding to $a_i$.

  The qubits from part b) are needed to ensure that a compute area of $L$ by $s$ is optimal. To ensure a width of at least $s$, one can devise a chain of CNOTs that have different control qubits but operate on the same target qubit. This is encoded in the qubits starting from the second and ending at qubit $(s+2)$ of part b). An additional column has to be created here, because one also has to ensure a height of $L$ in the compute area.

  \begin{figure}
    \centering
    \includegraphics[width=\columnwidth]{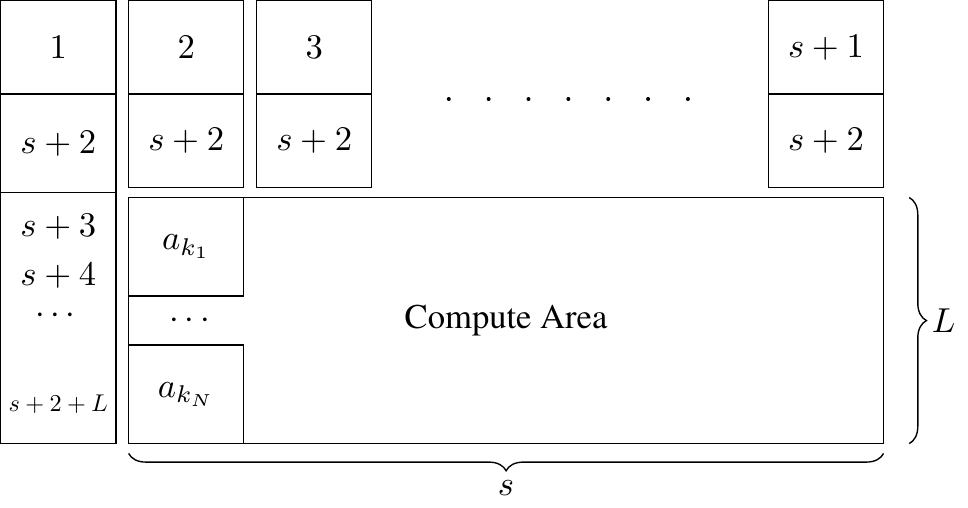}
    \caption{\label{fig:MappingLS} The circuit from Fig.~\ref{fig:MappingCirc} is now translated to the LS model of quantum computing. Here, the numbers indicate which qubit of the original circuit each patch represents. If the circuit can reach the theoretical optimum, the last $sL$ qubits can be fit in the compute area space. Each column then consists of $L$ patches of surface code, which are all filled with qubits that partake in CNOT operations. If each of these columns is completely filled, $s$ sets are found, which have elements that sum to $L$.}
  \end{figure}
  This can be performed by adding a single multiple target qubit CNOT with $L+1$ target qubits. One of these qubits is used to link the two helper CNOTs. This qubit is the $(s+2){\text{nd}}$ qubit in the circuit of Fig.~\ref{fig:MappingCirc}. The following $L$ qubits are used to increase the height by $L$. This results in the optimal placement of LS patches shown in Fig.~\ref{fig:MappingLS}. If that circuit cannot reach theoretical optimality, the compute area cannot contain all 3-partitioning CNOT-patches and thus additional qubits are needed. Holes (i.e., patches of surface code that do not correspond to any qubit in the circuit) are created and the bounding box of the calculation increases.

  The number of qubits that are needed for this mapping is $s \left(L+1\right) +L+2$. With the algorithmic ancillary patches, the circuit requires $\left(L+2\right)\left(s+1\right)$ patches in LS. Thus, this mapping only needs resources linearly in the number of integers of the original problem.

  By construction, each column in the compute area corresponds to one of the sets $A_i$, such that the requirement of each set summing to $L$ is equivalent to the requirement that each column in the compute area has a height of exactly $L$.
  Furthermore, checking whether each column exactly contains $L$ qubits can be performed in polynomial time, such that the problem is in the complexity class NP.

  The proof whether theoretical optimality is reachable implies that the optimization problem itself is NP-hard. This can be explained by using the optimization problem as a subroutine to the decision problem. If the optimization problem was easier, the decision problem would be solvable in polynomial time. With the described mapping, this would mean that any problem in NP could be solved polynomially, which is widely assumed to be false.

  \section{Discussion}
\begin{figure}
    \centering
    \includegraphics[width=\columnwidth]{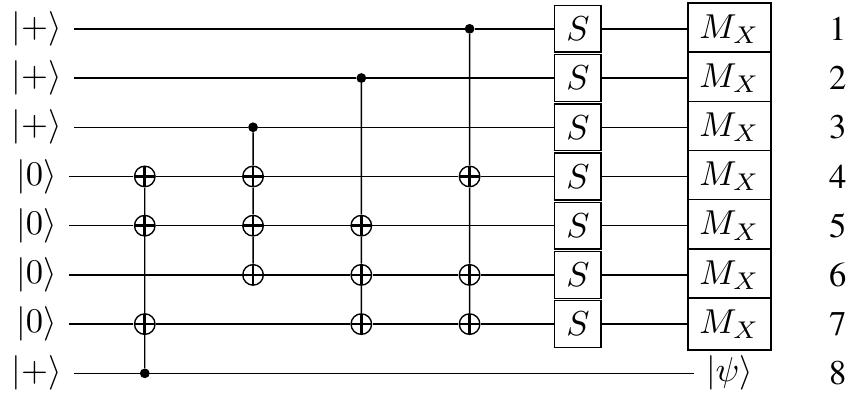}
    \caption{\label{fig:Steane} This circuit is the Steane code, to be used for the distillation of $\ket{Y}$ states. This is an iterative procedure where the error-prone $\ket{Y}$ are used during the application of the $S$-gates.}
  \end{figure}

  With this proof complete, we want to give an estimate on how much of an improvement can be expected from the optimization of a double $\ket{Y}$-state distillation circuit~\cite{Steane,Fowler2012}. This circuit is only illustrative for the optimization and we are aware of better proposals to implement $\ket{Y}$-states in surface codes~\cite{equivalence}. For reference, we provide the circuit of one distillation step in Figure~\ref{fig:Steane}. In our calculation we will give bounds for the best case by calculating the theoretical optimum. Furthermore, the worst case bounds are given by calculating an unoptimized placement, where each qubit corresponds to one row of patches in LS. However, previous manual optimization~\cite{herr_lattice_surgery} has shown that the $\ket{Y}$-state distillation circuit cannot reach theoretical optimality, such that the best possible solution lies somewhere in between these bounds. In the following back-of-the-envelope calculation we assume that the basis transformation of the measurement step can be applied without movement, which would correspond to a solution to a more complex optimization problem.

  The first round of the $\ket{Y}$-state distillation circuit consist of 7 distillations. Each distillation consists of 4 CNOTs from the Steane-code with 3 target qubits each. Furthermore, for the application of one $S$-gate an additional qubit is needed for the injection procedure. Thus, an additional $7 \cdot 7$ qubits are needed. In a second round, an additional distillation needs to be performed, requiring 8 more qubits and 4 more CNOTs. Furthermore, each distillation circuit consists of $8$ qubits, and initially $7 \cdot 7$ noisy $\ket{Y}$-states need to be injected. Thus, the total number of qubits needed in this double distillation is $N_Q = 8\cdot 7 + 7 \cdot 7 + 8 = 113$.

  The optimal costs can be calculated with Eq.~\ref{eq:cost} and lead to $32 \cdot 4 + 7 \cdot 7 = 177$ encoded patches of the planar code. A suboptimal placement (Eq.~\ref{eq:cost_unopt}), where each qubit is fixed to one row, requires $32 \cdot 113  = 3616$ patches. This difference is a factor of roughly $20$, with the difference only growing for larger circuits.
  
  Since this optimization has to be performed by a compiler, which is likely to run on classical hardware, the nature of the optimization being a NP-hard problem will restrict the size of exactly optimizable instance.
  To show that even a small amount of individual CNOTs would be unfeasible to exactly optimize, we look at an exact solution of the number-partitioning problem. An exact algorithm has to loop through all valid configurations to find the best one. We do not consider a dynamic programming solution here, because such an algorithm is unlikely to be devised for the optimization of LS.\@ The reason is that dynamic programming relies on the solution of subproblems. However, connections between different surface code patches required by merges break the structure exploited by dynamic programming for number partitioning. Furthermore, one should note that this optimization needs to be general, such that each circuit can be optimized. Any circuit-specific optimization is therefore discouraged, which makes our claims valid despite the symmetry of the current exemplary circuit.
  Assuming the same double $\ket{Y}$-state distillation circuit as before, we would have $46$ numbers and want to partition these into 15 subsets. The nature of the 3-partitioning problem only allows 3 numbers per set, such that we only have to consider these configurations. The assignment of $3N$ elements to $N$ sets such that each set contains exactly $3$ elements has
  \begin{equation}
    N_{\text{config}} = \prod_{i=0}^{N-1} \frac{\left(3N - 3i\right)!}{3! \left( 3N - 3i - 3\right)!} = \frac{\left(3 N\right)!}{\left(3!\right)^N}
  \end{equation}
  configurations. These would equal $\sim$$10^{44}$ possible configurations for the distillation circuit. A computer with 3.5 GHz and an ability to check one configuration per cycle would still need $\sim$$10^{34}$ processor hours to complete this task. Thus, it is not feasible to find the optimal solution with exact algorithms.
  The scaling of LS should be even worse, because individual qubits of the CNOTs have to be checked for eventual merges with horizontal neighbors adding an additional layer of complexity. Thus, this rough calculation indicates that the NP-hardness of this problem makes it impossible to optimize any meaningful quantum algorithms exactly and efficient heuristics have to be developed.

\section{Conclusion}
  We have proven that the decision problem of whether a circuit is perfectly optimizable using the LS-translation devised in~\cite{herr_lattice_surgery} is NP-complete and that the optimization problem has to be NP-hard. Furthermore, we have given some rough estimates on how hard exact optimization for LS would be, and showed that even small circuits cannot be optimized exactly. For practical purposes, however, the optimal configuration is not needed as an optimization protocol can get reasonably close.
  This urges further research in the development of efficient heuristics to optimize circuits in LS, that are as close to optimality as possible.
  Furthermore, the inclusion of the measurement step introduces an additional layer of complexity which has not yet been considered in our analysis. This will increase the space of possible configurations and likely decrease even further the efficiency of prospective optimization algorithms.

\begin{acknowledgments}
\section{Acknowledgments}We like to thank Michael J. Bremner and Austin G. Fowler for fruitful discussions and remarks. D.H. is supported by the RIKEN IPA program. S.J.D. acknowledges support from the JSPS Grant-in-aid for Challenging Exploratory Research and from the Australian Research Council Centre of Excellence in Engineered Quantum Systems EQUS (Project CE110001013).
S.J.D. and F.N. were supporeted by the IMPACT program of JST,
and the CREST grant No. JPMJCR1676.
F.N. was partially supported by the RIKEN iTHES Project,
MURI Center for Dynamic Magneto-Optics via the AFOSR Award No. FA9550-14-1-0040,
the Japan Society for the Promotion of Science (KAKENHI),
and the John Templeton Foundation.
\end{acknowledgments}

\end{document}